# Discovering the Computational Relevance of Brain Network Organization

Takuya Ito, Luke Hearne, Ravi Mill, Carrisa Cocuzza, Michael W. Cole*

Center for Molecular and Behavioral Neuroscience, Rutgers University, Newark, NJ, 07102, USA
*Correspondence: michael.cole@rutgers.edu (M.W. Cole).

## Abstract

Understanding neurocognitive computations will require not just localizing cognitive information distributed throughout the brain but also determining how that information got there. We review recent advances in linking empirical and simulated brain network organization with cognitive information processing. Building on these advances, we offer a new framework for understanding the role of connectivity in cognition – network coding (encoding/decoding) models. These models utilize connectivity to specify the transfer of information via neural activity flow processes, successfully predicting the formation of cognitive representations in empirical neural data. The success of these models supports the possibility that localized neural functions mechanistically emerge (are computed) from distributed activity flow processes that are specified primarily by connectivity patterns.

## Highlights

- Recent results suggest that localized functions in brain areas are specified primarily by their distributed global connectivity patterns
- Recent approaches go beyond stimulus/task brain mapping (and encoding/decoding) to characterize the role of brain connectivity in neural information processing
- We introduce network coding models as a framework encompassing neural network models optimized for task performance and those optimized for biological realism
- Biological and task performance constraints are complementary for aiding the search for accurate models of empirical brain function
- The activity flow algorithm is a core computational mechanism underlying network coding models, linking brain activity and connectivity in a biologically-plausible mechanistic framework

**Placing Brain Network Organization Within A Computational Framework**

A central goal of neuroscience is to understand how neural entities (such as brain regions) interact to compute cognitive functions. Historically, cognitive neuroscientists have tried to understand neural systems by mapping cognitive processes to neural entities. However, a comprehensive functional map of neural entities would still be insufficient to explain *how* cognition emerges through the collective interaction among these components. What could facilitate such mechanistic inferences? To answer this question, cognitive neuroscientists have begun to incorporate descriptions of brain network architecture (i.e. how neural entities are connected to each other) with these functional maps. Indeed, recent theoretical work suggests that mechanistic understanding of a brain system depends on characterizing the causal relationships between the system's components [1–3]. This is analogous to how we understand other mechanical systems; for example, understanding how a car moves upon pressing its accelerator requires knowledge of the causal relationships between the system's components (e.g., the pedal, the transmission, the engine, and the wheels). Similarly, mapping brain network connectivity allows for a mechanistic understanding by revealing the causal relations among neural components.

Facilitating this effort, recent advances in **network neuroscience** (see Glossary) are providing ever more detailed descriptions of brain network architecture [4]. Yet these descriptions have so far provided only limited insights into the neural computations underlying cognitive functions. Simultaneously, **connectionist** artificial neural network research has revealed the theoretical importance of network connectivity in computing cognitive functions (Box 1). Yet it remains unclear whether these theoretical insights emerging primarily from simulations apply to empirical neural computations. Thus, connectionism has provided an abundance of theoretical insights with limited translation to empirical evidence, while empirical neuroscience now has an abundance of empirical observations with limited theoretical insights. We propose that combining connectionist theory with empirical brain connectivity data will advance understanding of the computational and cognitive relevance of brain network organization.

We begin by discussing efforts to model neural computations using empirical brain connectivity. We then draw on conceptual insights from the connectionist literature – including from recent research involving "deep" neural network architectures – to help identify the computational and cognitive relevance of the distributed neural processes reported in recent empirical studies. Here we focus on a subset of studies that have successfully linked distributed brain connectivity with neurocognitive computations. This research is framed in a new data analytic perspective – **network coding models** – that suggests a way toward unifying empirical brain network analysis with a rich theoretical connectionist foundation.

**From Mapping Localized Functions to Building Network Coding Models**

Network coding models facilitate understanding of the function of localized neural entities (such as brain regions) by clarifying how they send and receive information: connectivity. This also has the advantage of clarifying the role of each brain connection in computing cognitive functions. In contrast to network coding models, cognitive neuroscience has primarily mapped tasks and stimuli to activity in neurons and neural populations – **function-structure mappings** (Figure 1A) [5]. Some examples of this general strategy include spike rate changes in single- or multi-unit recordings [6], general linear modeling with functional MRI (fMRI) [7], and event-related potentials with electroencephalography [8]. This strategy has been tremendously useful for characterizing the functions of spatially localized neural populations [9].

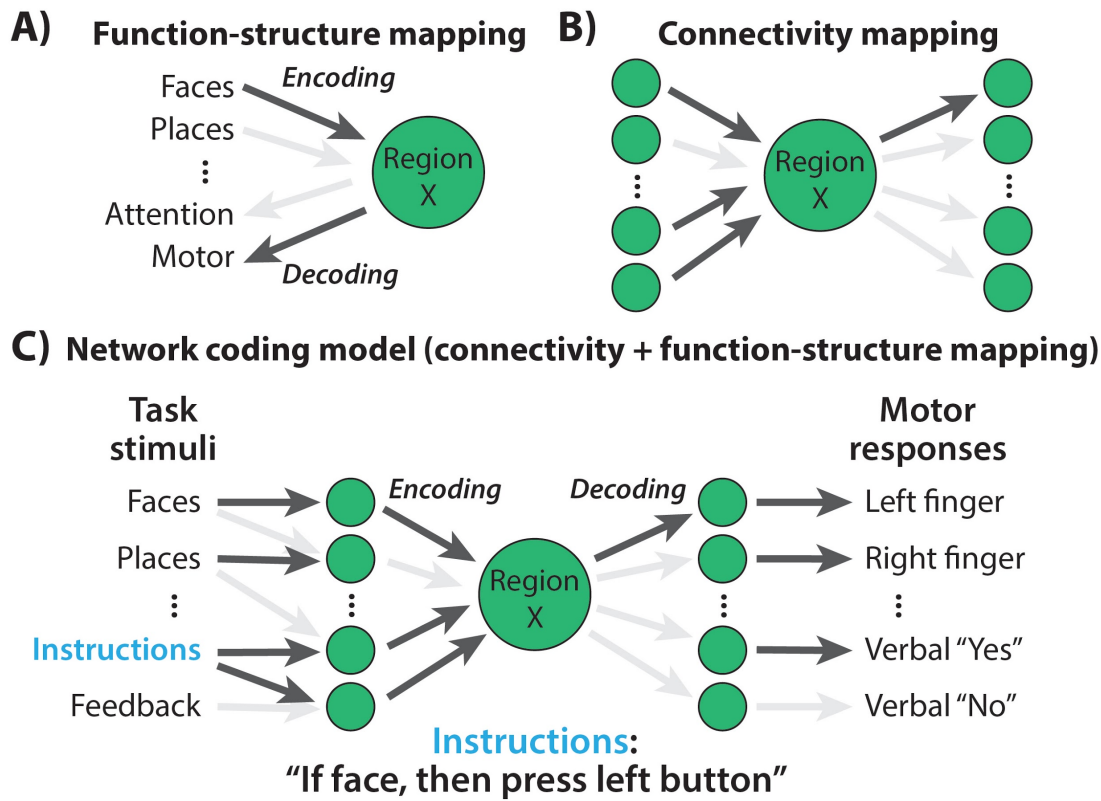

**Figure 1 – What would it take to understand a neural population's function? A**) The standard approach in neuroscience, identifying associations between cognitive variables and brain responses. This has led to sophisticated encoding (predicting neural activity from cognitive variables) and decoding (predicting cognitive variables from neural activity) models. Arrow darkness indicates strength of association between each cognitive variable and neural activity in Region X. Even with accurate associations it remains unclear *how* such selectivity arises in Region X. **B**) Knowing the set of connections to and from Region X provides additional mechanistic knowledge, yet it remains unclear what Region X represents or does in terms of information processing. Arrows indicate direction of causal inference (effective connectivity), but less detailed connectivity information (e.g., structural or functional connectivity) can also be useful. Arrow darkness indicates strength of connectivity between regions. **C**) We suggest that combining cognitive variables with connectivity in a single "network coding" model can allow for a more complete mechanistic understanding of a region's function. Connectivity can act as an encoding model for Region X, predicting its activity based on activity elsewhere in the brain. It can also act as a decoding model for Region X, predicting how Region X's activity influences other regions and ultimately behavior.

*Encoding and Decoding: Predictive Approaches for Mapping Localized Functions*

More recent methods have used a predictive framework to allow for more complex function-structure mappings. By predicting independent data, these approaches avoid overestimation of effects from increased model complexity (e.g., overfitting to noise), while providing more robust function-structure mappings [10,11]. It is commonly underappreciated how problematic overfitting is in standard data analysis approaches (e.g., general linear models or event-related potentials) that do not test for generalization in independent data using cross-validation [10,11]. Among these predictive approaches, an **encoding model** uses the task/stimulus condition to predict neural activity [12–14]. In contrast, a **decoding model** supports the opposite inference, using activity in a neural population to predict the task/stimulus condition [12–14]. These predictive models facilitate interpretation of neural activity in terms of task-related information content, which is an important step towards understanding the brain as an information processing system [15,16]. Notably, we have conceptualized network coding models within the framework of encoding and decoding models, since they inherit many of the benefits of encoding and decoding models while incorporating connectivity as a mechanistic **constraint** (as illustrated in Figure 1).

*The Role of Connectivity in Functional Localization*

Despite the utility of these approaches, a comprehensive function-structure mapping would not provide insight into how local functionality is specified mechanistically. As an example of mechanistic understanding, consider work in visual neuroscience suggesting that receptive field properties in V1 can be explained by connectivity patterns with the lateral geniculate nucleus [17]. More recent work has extended this insight to suggest that localized functionality in visual cortex might result from intrinsic connectivity architecture [18,19], with one study indicating that face and word selectivity in visual areas emerge from connectivity and specific computational processes [20].

Thus, one possibility to gain mechanistic insight is to map relationships between neural entities – **connectivity mapping** – based on the well-supported hypothesis that cognitive/behavioral functionality emerges due to neural interactions [21–24] (Figure 1B). Conceptually, this goal is similar to constructing a causal map among the functional components of a car (e.g., pedal -> engine -> transmission -> wheels). Various approaches to estimating connectivity could potentially be useful here, such as resting-state functional connectivity [25–27], task-state functional connectivity [28–34], or structural connectivity [35,36]. However, connectivity mapping can only facilitate mechanistic understanding to the extent that it can provide insight into the causal relations among neural entities [1,3]. Nonetheless, progress can likely be made even with limited causal inferences (e.g., structural/anatomical connectivity indicating possible rather than actual causal influence) by constraining the likelihood of possible causal models [1]. However, similar to the limitations of function-structure mappings, a complete connectivity map of neural populations (i.e., a **connectome**) would still not explain the emergence of any cognitive/behavioral functionality in a neural population [3,37]. Critically, unlike function-structure mappings, a connectivity mapping has no reference to cognitive/behavioral functionality. Without grounding in cognitive/behavioral functionality, a connectivity mapping merely describes potential routes of signaling among localized neural populations with no reference to the relevant cognitive/behavioral information content.

Therefore, we propose the importance of combining function-structure mapping and connectivity mapping to understand how function emerges from network connectivity in the form of network coding models (Figure 1C). The use of network coding models will facilitate a shift from purely functional models towards mechanistic models of brain function [15]. In the following section, we provide theoretical and empirical support for brain network organization as the basis for cognitive computations, highlighting converging ideas from network neuroscience and connectionist theory. We review instances of empirical studies that have bridged this gap, demonstrating how connectionist principles can be evaluated in models estimated from empirical neural data. We will conclude with a detailed description of network coding models (foreshadowed in Figure 1C) as a particularly useful approach to characterizing the role of brain connectivity patterns in cognitive computations. Ultimately, we expect these emerging areas of research to unify the major theoretical perspectives and methodological approaches in neuroscience: localized functions and information (reflected in task-evoked neural activity), and network neuroscience.

**Brain Network Organization Shapes Cognitive Computations**

*Connectionist Architectures and Cognitive Computations*

Decades of neuroscience have focused on function-structure mapping. Why might incorporating brain network connectivity (and, by extension, network coding models) address this strategy's limitations? Evidence from multiple sources suggests connectivity can provide a mechanistic explanation of how function emerges in neural networks (Box 1). First, decades of "connectionist" work with artificial **neural network models** have demonstrated the plausibility of distributed connectivity-based processes driving various complex cognitive functions [38–41]. Second, the standard model of neuroscience, as proposed by Ramón y Cajal [42] and solidified by Hodgkin and Huxley [43] and others [17], provides a prominent role for connectivity among discrete neural units in determining localized functionality. Third, there is increasing empirical evidence that the fundamental unit of functionality in the brain is not single neurons but rather populations of neurons [44–46].

In particular, neural network models have primarily utilized inter-unit connectivity to define the architecture for cognitive computations [47–49] (Figure 2A). This includes recently developed **deep neural networks** that improve model performance by including additional neural units with structured connectivity as "hidden" layers between input and output [47,49,50]. Thus, decades of modeling work demonstrates that connectivity architectures can support dozens (or hundreds) of complex cognitive processes, with more recent deep learning work indicating that additional performance gains are possible through refinement of connectivity architecture.

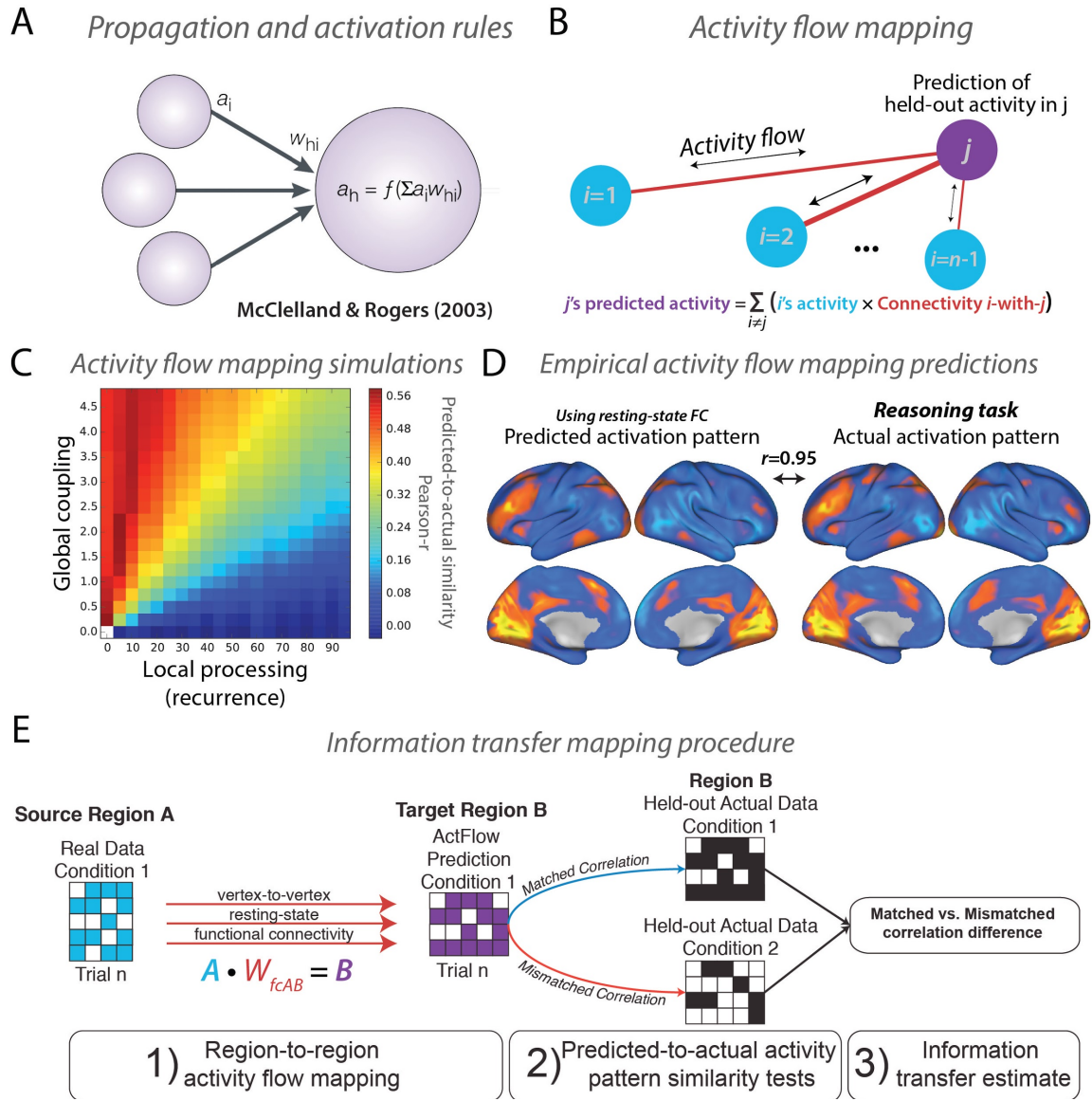

**Figure 2 – Activity flow as a linking principle between connectivity and activity. A**) Most neural network simulations use an algorithm to compute activity at each node at each time point. Briefly, each network node's activity ($a$) at the next time point is computed as the connectivity-weighted sum across the activity levels of all other nodes (the propagation rule), which is then passed through a function ($f$) (the activation rule) [51]. Figure adapted from [52]. **B**) Inspired by connectionist principles, the activity flow mapping algorithm implements basic neural network computations using empirical data, and accommodates limitations in empirical neural measures (such as limited fMRI spatio-temporal resolution). **C**) Simulations of fMRI data demonstrated that accurate activity flow mapping predictions depend on strong inter-unit connectivity (global coupling) relative to recurrent local connectivity. **D**) Activity flow-based predictions of task-evoked activity levels for a reasoning task (and many other tasks) were highly accurate, suggesting that empirical fMRI-based functional connectivity estimates are likely informative regarding the specification of localized functionality throughout the brain. Panels B-D adapted with permission from [53]. **E**) A modification of activity flow mapping to estimate information transfers, assessing whether decodable information in a source brain region can be preserved in a downstream target brain region. Panel E adapted with permission from [54].

These considerations support the conclusion that the brain's network organization is a major contributor to the computational architecture underlying its functionality, leading to a more focused question: how much function (in terms of both neural activity variance and cognitive/behavioral variance) can connectivity patterns explain? This is an important question since there are various alternatives to distributed connectivity in determining the functionality of single neurons and neural populations. For instance, relevant local neural population properties include population firing thresholds [43,55], neuron types [56,57], excitatory-inhibitory balance [58,59], local glial cell properties [60], and local recurrent connectivity [61,62]. Also highly relevant are neurotransmitter types (including large-scale neuromodulatory effects) [63,64] as well as systemic delivery of neuromodulators (e.g., hormones) [65].

*Testing Connectionist Principles in Empirical Data*

One way to evaluate the relative contribution of local versus distributed factors in determining the functionality of neural populations is to test the plausibility of each in computational models. We recently built on the fundamental algorithm underlying most computational simulations of large-scale neural processing – **activity flow** (Figure 2A) – to develop a procedure for testing local vs. distributed contributions to functionality [53,54,66] (Figure 2B). Activity flow is the movement of neural activity (e.g., spike rates or fMRI signal amplitudes in biological contexts) between neural entities. Quantitatively, the activity flow approach involves simulating a single neural network computation (e.g., a forward iteration in a feedforward neural network): $a_h = f(\sum_{i \in I} a_i w_{hi})$, where a unit $a_h$'s activity is a linear combination of all other units activity ($\sum_{i \in I} a_i$) weighted by their connectivity ($w_{hi}$) to $a_h$ before passing through a transfer function $f$, such as a sigmoid. Within the connectionism framework, this formalization is a combination of the propagation and activation rules [51]. Other names given to more specific instantiations of this algorithm (e.g., with a particular form of nonlinearity) include the McCulloch-Pitts neuron [55], the perceptron [67], the divisive normalization model of neural activity [68], adaptive linear element [69], and spreading activation [70]. Critically, this algorithm was adapted for use with empirical data (e.g., fMRI activity and functional connectivity estimates) to parameterize empirically-derived models that make quantitative predictions of the spread of estimated activity over brain network connections (Figure 2B).

In neural network simulations we found that activity flow mapping was only effective in network architectures with relatively large effects of distributed (relative to recurrent/local) connectivity (Figure 2C). This is consistent with previous findings, where large effects of inter-regional synaptic coupling (relative to local coupling) were important for predicting functional from structural connectivity [71]. Thus, when we applied this approach successfully in empirical fMRI data across a variety of diverse tasks (see Figure 2D for an example) we were able to conclude that distributed connectivity plays a substantial role in determining localized functionality [76]. This is highly compatible with the notion that each localized population has a "connectivity fingerprint" that largely determines its functionality [72,73]. These results are also in line with the observation that large-scale propagation of neural activity in animal models tends to conform to large-scale anatomical connectivity patterns [74].

Other studies have further demonstrated the role of brain connectivity in mediating distributed neural activity. One study found that decodable cognitive information transfer between pairs of brain regions can be mediated via activity flow processes [54]. This illustrated that information in a source brain region can be sent (encoded) through network connectivity to be received (decoded) by a target brain region (Figure 2E). In another study, we found that the task activations of healthy older adults can be transformed into dysfunctional task activations of unhealthy older adults via the latter's connectivity patterns [66]. This was accomplished by estimating activity flow through disrupted intrinsic functional networks of subjects at increased risk for developing Alzheimer's disease. This suggests that changes in brain network organization underlie "unhealthy" cognitive computations [66].

*Using Connectionist Models to Understand Empirical Data*

A general principle of connectionist theory is that the ability of a neural network model to perform an experimental task is dependent upon its connectivity patterns [46]. This suggests an alternative approach to study the importance of connectivity in producing cognitive computations: training an artificial neural network model to perform an experimental cognitive task. Despite the connectivity patterns of these models being specified more by task training than biological constraints, many of these models are able to accurately predict empirical neural responses [40,41,46,75]. Moreover, both biological and artificial systems can be characterized by the same computational mechanism: activity flow processes (Figure 2A) [52]. This again supports the conclusion that connectivity patterns – even those derived primarily from task training constraints – are central to neural computations underlying cognitive processing.

Despite these promising results, much work remains to determine the role of the brain's network architecture in influencing cognitive functionality. For instance, the particular cognitive tasks used are only a small sample of the wide array of tasks humans are capable of [76,77]. Thus, much work remains to verify the role of connectivity patterns in determining the computations across diverse tasks and stimuli. Notably, understanding how many different tasks can be encoded through distributed network processes will likely facilitate the design of generalized models of cognition that can adaptively perform novel tasks [13,76,77]. Additionally, it will be important to assess how activity flow computations are altered across different levels of functional organization, from finer-grained activity and connectivity patterns to large-scale functional brain regions and networks. Finally, it will be important to determine the role of features other than connectivity – such as local activation functions (incorporating nonlinearities, such as compressive summation in visual cortex) [78,79] and operating timescales/dynamics [71,80] – in specifying the neural computations that facilitate cognitive processes.

**Additional Approaches for Mapping Cognitive Function with Connectivity**

Given the strong evidence that connectivity is central to neural computation, any methods that link cognitive function with connectivity are likely to provide useful theoretical insight. In this section we focus on efforts that characterize information in distributed networks, as well as efforts that quantify how cognitive information representations quantitatively change between brain areas. In the subsequent section we will focus on network coding models, which provide a mechanistic characterization of information processing, due to the incorporation of connectivity estimates as constraints. For instance, there have been significant advances in characterizing how the synchronization of neural time series during tasks – task-state functional connectivity – relates to ongoing cognitive processes. While links between resting-state functional connectivity and cognitive ability (estimated via individual difference correlations) have been widely reported [21,22,81–83], changes to the underlying functional network organization from resting state to task state can shed light on how those network changes contribute to task-related cognitive processes [30,34].

Recent work has shown that the functional network organization during resting and task states are highly similar [29,31], with the functional network organization at rest accounting for up to 80% of the variance in whole-brain network organization during tasks. However, studies have reported systematic task-related changes in functional network organization that reflect ongoing cognitive processes [28,33,84]. Moreover, transient changes in task-state functional connectivity have been shown to predict task performance [84]. Thus, measures of task-state functional connectivity can provide insight into which brain region interactions are involved in a cognitive process.

Despite the insights offered by standard task-state functional connectivity analyses, correlation-based measures of the BOLD signal between pairs of regions limit the identification of what *kind* of information is transmitted between brain regions and *how* this information might be transmitted. Several recent approaches have gone beyond standard functional connectivity measures, providing multivariate measures of temporal and spatial dependence between brain areas [85]. For example, one study measured the time-varying information content (i.e., decodability) during task states, and correlated the informational time series with other brain regions [86]. This technique goes beyond asking whether two brain regions are synchronized and addresses whether pairs of regions contain task-related information at the same time.

Other approaches estimate the neural transformation of cognitive representations between brain regions, clarifying the mathematical relationship between distinct units. One approach maps spatial activation patterns between brain areas using a nonlinear transformation, capturing the optimal or "normative" computational transformation required to project one brain region's information into another brain region's geometry [87]. Another similar approach estimated the optimal linear transformation required to project activation patterns in early visual cortex to regions further along the ventral visual stream, such as the fusiform face complex [88]. By identifying a simple linear transformation, the authors were able to investigate the computational properties of the linear transformation matrix, such as whether the mapping projected to a lower dimensional space (e.g., information compression) as the information was mapped from early visual areas to the fusiform face complex. Thus, by characterizing the representational mappings between brain regions during cognitive tasks, these studies go beyond what standard task-state functional connectivity approaches offer to characterize inter-unit computational relationships. However, such models fail to constrain their predictions with separate estimates of brain connectivity; a key aspect of network coding models.

**Network Coding Models: Computing Cognitive Information in Neural Networks**

We and others have made the case that a particularly powerful framework for characterizing the functionality of brain regions is to use encoding and decoding models [12–14]. However, most uses of encoding and decoding models are designed to characterize information of interest to the experimenter, and are inconsistent with how neural entities likely encode and decode task information biophysically [89,90]. An example of this are function-structure mappings that map high-level, human annotated information (such as semantic information) onto brain activity [91]. Though these data-driven functional models can be useful, neural entities likely encode and decode complex task features through network connectivity, rather than the direct task stimulus-to-neural response associations composing traditional encoding and decoding models (Figure 3A). The absence of mechanistic constraints (such as network connectivity) in these models limit the causal relevance of experimenter-extracted neural representations [89,90].

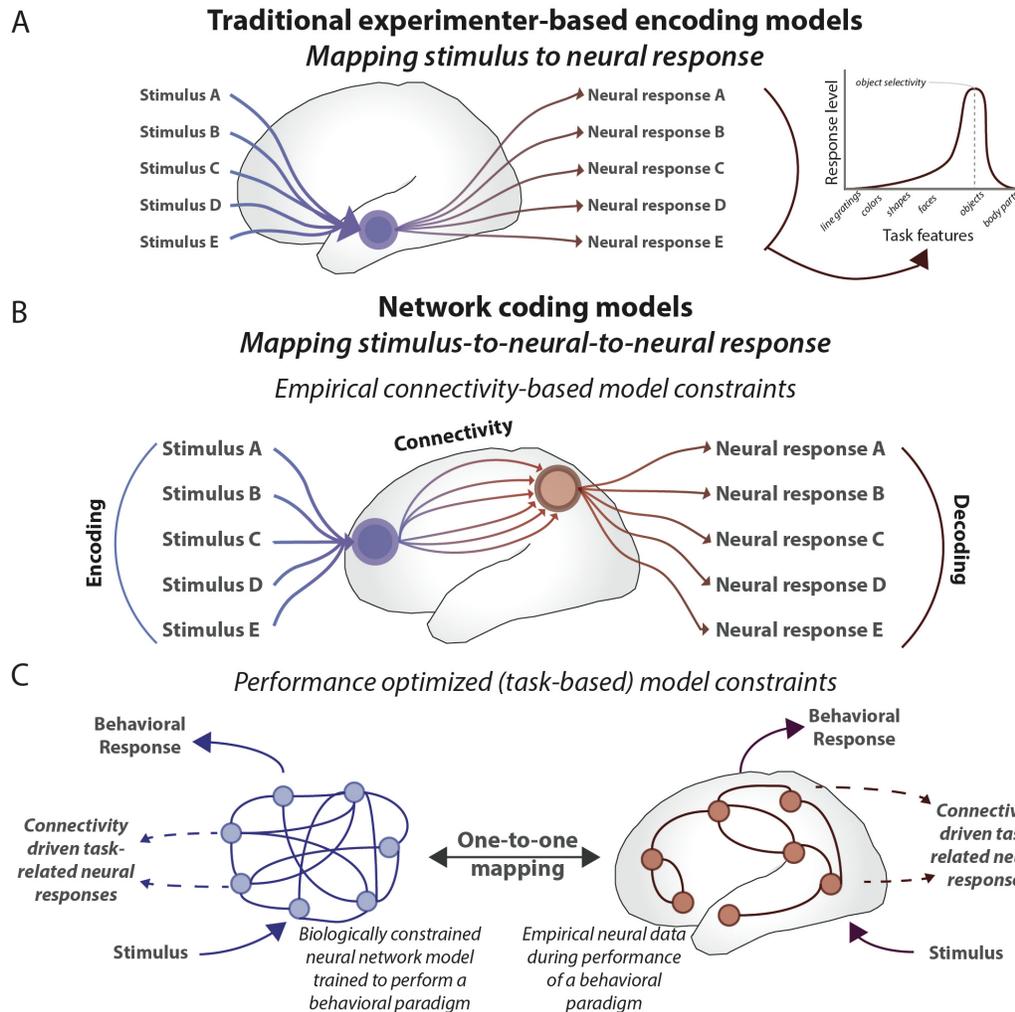

**Figure 3. From experimenter-based encoding/decoding models to network coding models. A)** Traditional encoding models impose experimenter-based stimulus constraints to measure neural responses. These measures represent the degree to which a neural entity responds to different task features. Despite the simplicity of these models, the neural responses predicted by these encoding models may not reflect how the brain actually responds to the task information. Instead, these models focus on whether the experimenter can map task information onto neural responses, rather than how a neural entity actually encodes task information through its network connectivity [89]. **B)** Network coding models (see Figures 1 and 2) utilize mechanistic constraints (e.g., brain network connectivity) to investigate how a neural entity's connections might drive its task-related response. The first approach is to estimate connectivity directly, and then predict (via activity flow estimation) neural responses in a downstream neural entity to characterize how those responses likely emerge from activity flow processes via connectivity. **C)** The second, more top-down approach (see Figure 4) uses behavior or task performance to learn the network connectivity patterns necessary to encode task representations (through learning algorithms). These artificial neural networks implicitly model activity flow processes to implement the neural computations underlying cognitive task functions. When provided with sufficient biological constraints (e.g., the visual system's "deep" network architecture), these models can be directly compared to empirical data, producing unique insight into the network principles that are instrumental to performing cognitive processes (which include cognitive encoding and decoding) [41,46,48].

A potential solution to this problem is to incorporate network connectivity estimates in conjunction with traditional encoding and decoding models – a network coding model (Figures 1C, 3B, and 3C). Figure 3 formalizes the distinction between standard encoding/decoding models and network coding models. The latter approach would add mechanistic constraints – brain network connectivity – to encoding and decoding models. Constraining models via connectivity, as performed in activity flow mapping (Figure 2), enables a stronger mechanistic interpretation of how neural entities receive (encode) and send (decode) information within the brain. Additionally, this approach explicitly implements activity flow processes in empirical data (Figure 2A). This increases the biological relevance of these models by simulating the physical process by which neural signals are relayed through network connectivity [53,54,66]. Ultimately, this framework strengthens causal inferences made about the transmission of task information within the brain, beyond traditional descriptive statistics [3,54].

Network coding models provide a biologically plausible link between encoding/decoding models and connectionist theory when studying the propagation of activity during cognitive tasks. Several recent methods have applied this general framework. The first is to incorporate estimates of functional connectivity directly with traditional, **experimenter-based encoding and decoding models** (Figure 3B) [54]. This approach tests whether decodable information in one brain area can be re-encoded through connectivity patterns and decoded in a downstream target brain area, providing a biophysically plausible model of information transfer. We recently demonstrated that task information in a set of brain regions could be used to predict the same task features in downstream regions through activity flow processes using functional connectivity estimated using resting-state fMRI (Figure 2E) [54]. These findings illustrate that task information in a brain area not only can be encoded/decoded by the experimenter, but is also used by other brain areas through distributed network connectivity.

The second approach uses **structured connectionist models** (neural network models with architectural constraints) to study the emergence of localized functionality from connectivity (Figure 3C). Recent technological advances in the training of structured connectionist models, such as biologically-inspired deep neural networks and **recurrent neural networks**, have enabled the study of the encoding and decoding of cognitive information via activity flow through optimized connectivity architectures. For example, a recent study showed that a recurrent neural network was able to represent an array of different inputs, such as different task states and stimuli (encoding), and map those inputs to different motor outputs (decoding) for accurate task performance [41]. Importantly, the task features/stimuli that this neural network encoded (its inputs) were qualitatively different from the features that it decoded (motor outputs), offering insight into how information might be transformed across neural entities via network connectivity [49]. Thus, while traditional experimenter-based encoding and decoding models typically address *what* a neural entity might be encoding and decoding, network coding models can also address *how* that information might be computed and/or subsequently processed.

Although unconstrained/unstructured neural network models are universal function approximators [92], employing biological priors in neural network models can aid in producing mechanistic models of neural computations (Figure 3C) [93]. Such constraints (model assumptions) are essential to discovering the model features that are involved in the neural computations underlying cognitive functions. For instance, connectivity architecture and task performance constraints were implemented to demonstrate how images can be encoded and decoded to identify objects in neural network models in both human [50] (Figure 4) and non-human primates [47,48]. These types of models map visual images onto specific object codes via a nonlinear computation mediated by connectivity (Figure 4A**)**. While any neural network model can be theoretically trained to accurately perform this computation, Wen and colleagues demonstrated that when adding biological constraints into this neural network (e.g., number of layers and number of units per layer to match those of the ventral visual stream) the network exhibited similar neuronal responses to empirical fMRI data obtained in humans (Figure 4B) [50]. Moreover, they identified a face-selective unit in their model and showed that its activity was highly correlated with the fusiform face area in empirical fMRI data when dynamic naturalistic stimuli was shown to both human and model (Figure 4C). This suggests that like the brain, localized functionality can emerge in neural network models even when primarily optimized for task performance. Critically, incorporating realistic biological constraints can improve the fidelity with which artificial neural networks mimic the brain in representing cognitive information.

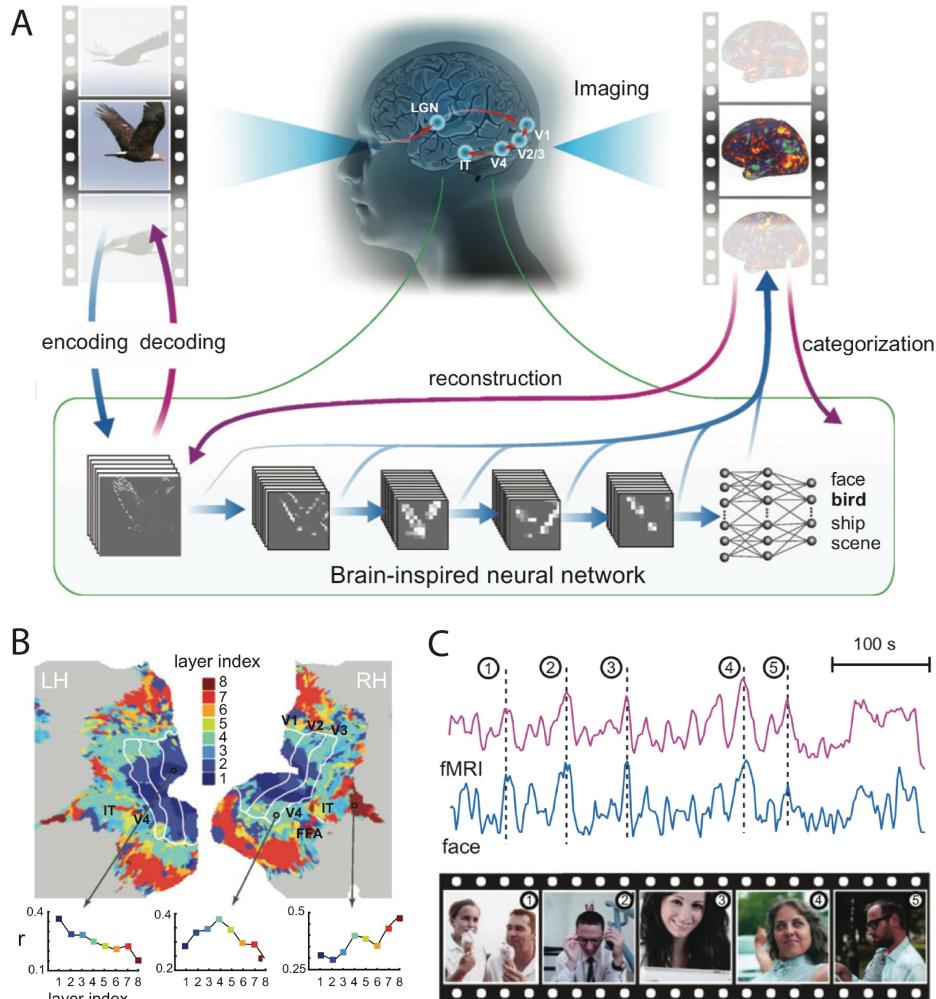

**Figure 4. Leveraging neural network models to explain empirical neural data in the visual system.**
**A)** When appropriately constrained with biological and structural priors, artificial neural networks can be used to model empirical neural data [50]. Wen and colleagues compared representations in a brain-inspired neural network to representations found in empirical fMRI data during presentation of the same naturalistic stimuli (movies). **B)** Stimulus activations in different layers of the deep convolutional neural network corresponded to different brain areas in the visual system. For example, brain areas in earlier visual processing areas (e.g., V1 and V2) contained representations more similar to representations in earlier layers in the artificial neural network. **C)** The activation time series of face-selective areas in fMRI data and the face-selective unit in the artificial neural network had highly similar and comparable time series during presentation of the dynamic naturalistic stimuli. Figures adapted with permission from [50].

Network coding models can therefore provide directly testable hypotheses in empirical data. For example, if a neural network model contains enough biological constraints, a mapping between the network model and empirical neural data can be made (Figure 3C). In a recent study, a biologically-constrained artificial neural network of the ventral visual stream provided insight into how to control and activate targeted neural populations in area V4 [75]. By dissecting the receptive fields of specific units in their biologically-constrained artificial neural network model, the authors generated novel visual stimuli that could theoretically target a subset of biological units in the primate V4. Indeed, the authors showed that these stimuli could activate and control corresponding neural populations empirically in the primate brain, uncovering the receptive fields (and localized function) of those biological units through model analysis. This approach provides a powerful framework to test how localized functionality emerges from network organization in both artificial and biophysical neural networks. We suggest that including additional biological constraints, such as empirically-derived connectivity estimates, would likely lend additional mechanistic insight into how neural information processing is carried out in the brain [15].

**Concluding Remarks**

The recent proliferation of large neural data sets with rich task features has created a wealth of opportunities in functional brain mapping. However, data-driven approaches to mapping task features to neural responses largely disregard the biological mechanisms of neural information processing: distributed cognitive processing through brain network connectivity. Since the early days of connectionism, the functionality of neural entities has long been hypothesized to be embedded in its patterns of network connectivity [51,72,73]. Here we propose the incorporation of brain network connectivity as a biological constraint underlying the emergence of localized functionality and distributed cognitive processing.

Though mounting evidence suggests that connectivity undoubtedly plays a role in cognition, the precise contribution of distributed versus local processing remains an open question (see Outstanding Questions). Several advances are needed to answer this question. First, it will be important to advance current connectivity estimates, given that current methods may only indicate associational (and not causal) relations [3,94]. Second, it will be important to integrate additional biological properties that shape local processing properties (such as nonlinear transfer functions) and incorporate them into network coding models [20]. Lastly, it will be important to evaluate how these principles might generalize across levels of organization and spatial scales using the wide variety of data types available to neuroscientists. Such insight can provide more mechanistic insight into how local functionality emerges from the interplay between distributed (e.g., connectivity) and local (e.g., nonlinear) processes. Taken together, we can begin to reconstruct how the collective interaction of functionally-specific localized processes work together to compute diverse cognitive functions.

**Box 1. The evolution of connectionism in relation to neuroscience**

Despite the biological and neural origins of connectionism [38,55,67,69], there are stark differences in current connectionist (deep learning) and neuroscientific research. One of the most influential advances in connectionism was the discovery of backpropagation, an algorithm that enabled the training of multilayer artificial neural networks. Pioneered by Paul Werbos in 1974 and popularized by other connectionist researchers [51,69,95], backpropagation provided researchers with the tools to train networks to perform increasingly complex tasks. This discovery strongly influenced the direction of connectionist research away from its neurobiological roots and towards the development of learning algorithms. Until recently, the development of these learning algorithms has been the primary focus of many connectionist and deep learning researchers.

Despite this focus on learning algorithms, learning in artificial systems is in many ways incomparable to learning in biological systems. One reason for this is the biological implausibility of the backpropagation learning algorithm, though biologically-plausible equivalents exist [96]. A more fundamental reason is that biological systems are born with inductive biases (i.e., innate structure) that shape how they interact with their environments early in life [93,97]. A recent opinion paper argued that the inductive biases afforded to a biological system are the result of learning during evolution (rather than within a single lifespan), and is likely embedded in an organism's DNA [97]. Thus, learning in artificial systems is more appropriately compared to learning on evolutionary timescales (which often takes place outside the context of neuroscience research), since artificial neural networks are typically "born" with a blank slate.

The network coding model approach advocated here is consistent with recent work focused on integrating inductive biases into neural network models to understand neurocomputational principles. Such biologically-based inductive biases, such as the incorporation of highly structured network architectures, have been especially useful in understanding core object recognition in the ventral visual stream [47,48,50,75,98]. Other inductive biases, such as the incorporation of cortical heterogeneity and local biophysical properties, have been especially useful in characterizing simulated large-scale network dynamics [62,99–101]. Here we focus primarily on the incorporation of empirical brain network architecture to provide principled inductive biases to predictive (encoding/decoding) brain models. We expect insights from these network coding models to positively impact both neuroscience and connectionist research.

**Glossary**

**Activity flow:** A fundamental computation (see Figure 2A) describing the movement of activity between neural units as a function of their connectivity (e.g., propagating spikes over axons in the brain). This is equivalent to the activation and propagation rules used in connectionist research.

**Connectionist/Connectionism:** A subfield within cognitive science that focuses on implementing cognitive and mental phenomena computationally through artificial neural networks.

**Connectivity mapping:** The quantification of the relationship between two or more neural entities (e.g., brain regions) using either statistical dependencies of the neural time series (functional connectivity) or estimates of structural/anatomical pathways (structural connectivity).

**Connectome:** A complete connectivity map of all units in a neural system.

**Constraint:** A limitation placed on a model in an optimization context. We focus primarily on empirical constraints, especially biological constraints such as brain network organization. We also focus on task performance constraints, which are typically emphasized in connectionist modeling.

**Decoding model:** A statistical model that predicts a task stimulus or condition as a function of a set of neural responses.

**Deep neural network models:** Neural network models with more than one hidden layer, which have been shown to boost task performance in many cases relative to traditional neural network models.

**Encoding model:** A statistical model that predicts a neural response as a function of a task stimulus or condition.

**Experimenter-based encoding/decoding models:** In contrast to network coding models, encoding/decoding models that focus on how the experimenter encodes/decodes information from a neural entity, rather than how other neural entities in the brain encode/decode that information.

**Function-structure mapping:** The association between a particular neural entity and its functionality, such as what task stimuli a neural entity activates or responds to.

**Network coding models:** Models of brain function that simulate encoding and decoding processes through network connectivity (typically via the activity flow algorithm) to predict empirical brain activity.

**Network neuroscience:** A subfield of neuroscience concerned with understanding the network principles of neural phenomena by exploiting tools from network science.

**Neural network models:** Computational models consisting of a network of interconnected units that are optimized to match biological features (biological constraints) and task performance (normative task performance constraints) to varying extents.

**Recurrent neural network model:** Neural network models that feedforward through time, rather than through different spatial layers. Recurrent connections indicate directed connections to other units within the network that propagate activity through time.

**Structured connectionist models:** Neural network models with built-in architectures (i.e., innate structure), such as many deep neural networks with a fixed number of layers.

**Acknowledgements**

The authors would like to thank Ruben Sanchez-Romero for assistance with preparing this article. The authors acknowledge the support of the US National Institutes of Health under awards R01 AG055556 and R01 MH109520, and the Behavioral and Neural Sciences Graduate Program at Rutgers, The State University of New Jersey. The content is solely the responsibility of the authors and does not necessarily represent the official views of any of the funding agencies.